\documentclass{article}

\usepackage[english]{babel}
\usepackage{rotating}
\usepackage{placeins}
\usepackage{float}
\usepackage[a4paper,top=2cm,bottom=2cm,left=3cm,right=3cm,marginparwidth=1.75cm]{geometry}
\usepackage{blindtext}
\usepackage{amsmath,enumerate,array}
\usepackage{graphicx}
\usepackage[colorlinks=true, allcolors=blue]{hyperref}
\usepackage{rotating}
\usepackage{graphicx}
\usepackage{lipsum}

\makeatletter
\renewcommand{\maketitle}{\bgroup\setlength{\parindent}{0pt}
\begin{flushleft}
  \textbf{\@title}

  \@author
\end{flushleft}\egroup
}
\makeatother

\title{\LARGE \textbf{A mathematical model of thyroid disease response to radiotherapy}

\bigskip}

\author{Araceli Gago-Arias$^{1, 2, 3}$*, Sara Neira $^{1}$, Filippo Terragni$^{4}$ and Juan Pardo-Montero$^{1,2}$* \\

\bigskip

\small ${1. }$Group of Medical Physics and Biomathematics, Instituto de Investigaci\'{o}n Sanitaria de Santiago (IDIS), Santiago de Compostela, Spain.\\
\medskip
\small  ${2. }$Department of Medical Physics, Complexo Hospitalario Universitario de Santiago de Compostela, Spain.\\
\medskip
\small   ${3. }$Institute of Physics, Pontificia Universidad Cat\'{o}lica de Chile, Santiago de Chile, Chile.\\
\medskip
\small  ${4. }$G. Mill\'{a}n Institute and Department of Mathematics, Universidad Carlos III de Madrid, Legan\'{e}s, Spain.\\

\bigskip \textbf{*Corresponding author:} Araceli Gago-Arias, E-mail: maria.araceli.gago.arias@sergas.es; and Juan Pardo-Montero, E-mail: juan.pardo.montero@sergas.es \\ }

\begin{document}

\maketitle

\begin{abstract}
We present a mechanistic biomathematical model of molecular radiotherapy of thyroid disease. The general model consists of a set of differential equations describing the dynamics of different populations of thyroid cells with varying degrees of damage caused by radiotherapy (undamaged cells, sub-lethally damaged cells, doomed cells, and dead cells), as well as the dynamics of thyroglobulin and antithyroglobulin autoantibodies, which are important surrogates of treatment response. The model is presented in two flavours: on the one hand, as a deterministic continuous model, which is useful to fit populational data, and on the other hand, as a stochastic Markov model, which is particularly useful to investigate tumor control probabilities and treatment individualization. The model was used to fit the response dynamics (tumor/thyroid volumes, thyroglobulin and antithyroglobulin autoantibodies) observed in experimental studies of thyroid cancer and Graves' disease treated with $^{131}$I-radiotherapy. A qualitative adequate fitting of the model to the experimental data was achieved. We also used the model to investigate treatment individualization strategies for differentiated thyroid cancer, aiming to improve the tumor control probability. We found that simple individualization strategies based on the absorbed dose in the tumor and tumor radiosensitivity (which are both magnitudes that can potentially be individually determined for every patient) can lead to an important raise of tumor control probabilities.\\

\end{abstract}

\section{Introduction}
\label{sec_intro}
Radioactive iodine, $^{131}$I, therapy (RAI) is a type of molecular radiotherapy (also called targeted radiotherapy) that has been commonly used for the treatment of differentiated thyroid cancer since the 1940s. This therapy, based on the ability of well-differentiated (papillary or follicular) thyroid cancer cells to absorb iodine, uses radiation either to ablate remnant thyroid tissue after surgery or to treat residual, recurrent, or metastatic cancer, such as pulmonary and bone metastases. Radioiodine therapy is suitable for patients with I-avid but surgically unresectable metastases of differentiated thyroid cancer, DTC. In patients with non-avid metastases, due to the expansion of less differentiated cells, RAI shows no benefits, and treatment shifts to other strategies.

One of the most relevant physical quantities affecting the response to therapy is the absorbed radiation dose to lesions~\cite{maxon1983,jentzen2014,wierts2016}. Large differences in the efficacy and cure rates of RAI on metastatic DTC patients are reported in many studies~\cite{schlumberger1996}. This might be partially due to the large dosimetric uncertainties of RAI. Other factors, such as age, thyroglobulin levels at diagnosis, and stage of the disease have been identified as important prognostic factors~\cite{wang2017, gleisner2017}. The systemic nature of this therapy, with a radionuclide that takes several days to disintegrate, makes RAI treatment planning a complex procedure. Due to these complexities, RAI treatment administration is usually guided by non-personalized empirical criteria that establish the activity of $^{131}$I to be administered, as well as cycle schedules. This scenario poses a concern of potentially over/under-dosing patients, and there is a need to develop alternative strategies and tools for the development of individualized treatment planning.

Some markers are usually measured in the management of thyroid cancer treatment, as surrogates of response. The most extended tumor marker is thyroglobulin, Tg, which is produced by normal follicular cells, and accumulated as a colloid in an extracellular compartment of the thyroid follicles. Under certain circumstances, such as rapid and disordered thyroid tissue growth, inflammation leading to follicular destruction, or hemorrhage, Tg is also secreted into the systemic circulation. The concentration of Tg in blood is thus proportional to the volume of thyroid tissue, and elevated serum Tg levels can be seen in benign conditions like Graves’ disease and thyroiditis, and also in thyroid cancer. Regarding the latter, high serum Tg concentrations are observed in approximately two-thirds of well-differentiated thyroid cancers, including follicular and papillary thyroid carcinomas, up to 50\% of anaplastic thyroid carcinomas, and some medullary carcinomas~\cite{indrasena2017}.

After partial or total thyroidectomy, serum Tg levels post-surgery reflect the amount of residual thyroid mass, with high levels one month after thyroidectomy indicating incomplete excision of the gland, or recurrent or metastatic disease~\cite{giovanella2010,pelttari2010}. Furthermore, the risk of having recurrent or persistent disease increases as the postoperative Tg rises~\cite{piccardo2013}. This is used in the initial follow-up of patients with papillary and follicular thyroid carcinoma, with measurements every 6--12 months, as a guide to initiate RAI for remnant ablation, and as an important surrogate marker of the response of metastatic patients to RAI~\cite{haugen2016}. However, the subset of DTC patients (12\%) with low blood Tg levels before thyroidectomy will not show rising serum Tg levels when there is a recurrence of the disease, therefore the importance of knowing the patient serum Tg level before surgery~\cite{brendel1990}.

The measurement of serum Tg is influenced by the coexistence with antithyroglobulin autoantibodies, TgAb, in blood. These antibodies are present in cases of Hashimoto’s disease, Graves’ disease, 7.5--25\% of DTC patients and 10\% of the healthy population~\cite{spencer1998}. Their presence impairs the reliability of the most sensitive Tg detection methods currently available (immunometric assays, IMA), leading to levels lower than expected or undetectable~\cite{grebe2009,netzel2015}. Therefore, current thyroid cancer management guidelines indicate that TgAb should always be measured in parallel with Tg, to identify potentially unreliable Tg tests~\cite{haugen2016}. The analytical interference is usually attributed to the ability of serum TgAb to block the access to the Tg epitopes, to which the IMA test antibodies bind~\cite{indrasena2017,spencer2014}. In addition, it has been recently reported that the interference is more significant in vivo than in vitro, suggesting the induction of the enhanced clearance of Tg by TgAb~\cite{latrofa2016,latrofa2018,ricci2020}.  Since the only source of Tg is thyroid tissue, the persistence of elevated TgAb levels is an indicator of the continued presence of Tg, and, for example, possible metastasis. As TgAb levels change in accordance with Tg concentration, follow-up measurements of these antibodies can be used as a surrogate tumor marker, helping to predict cancer recurrence in patients with disseminated thyroid carcinoma~\cite{jo2018,sinclair2006}.

There is an ongoing paradigm shift towards individualized treatment planning in molecular radiotherapy, including RAI. To achieve this goal, there is a need to develop individualized dosimetry methods for molecular radiotherapy~\cite{lassmann2010}, but also reliable biomathematical dose-response models that can predict the effect of the treatment~\cite{barbolosi2017,traino2006}. Mathematical models can be very useful to assist in the analysis and interpretation of experimental results. Once validated, they can be used to investigate “What if?” questions, simulating different treatment strategies to optimize RAI through computational simulation. The use of computer modeling is usually referred to as {in silico} modeling, as an allusion to the {in vivo} or {in vitro} models typically used in biology/biomedicine. Such models might also be useful to guide the management of patients during the course of the treatment, especially if they include the markers usually measured in the clinical practice as surrogates of response. In the case of the thyroid, models of response to cancer (or other diseases) therapy, including RAI, should ideally include the evolution of Tg and TgAb. However, this has not yet been addressed in the literature.

Modeling of thyroid cancer response to therapy was previously studied by Barbolosi~et~al.~\cite{barbolosi2017}, who presented a phenomenological model of tumor response to RAI, including dynamics of Tg. Traino~et~al.~\cite{traino2006} provided a differentiated thyroid cancer mass reduction model, based on personalized RAI dosimetry calculations, for post-surgical thyroid remnant or recurrent or metastatic cancer. That study did not consider the evolution of Tg or TgAb. Pandiyan~et~al.~\cite{pandiyan2018} presented a mathematical model of response to methimazole therapy for Graves’ disease, which includes many of the compartments discussed above, and models the evolution of TSH, T4 and T3 hormones, as well as other antibodies relevant for Graves' hyperthyroidism, but not Tg nor TgAb. Other mathematical models of Graves' disease relapse after antithyroid drugs~\cite{langenstein2016} or RAI~\cite{traino2005} have been also presented. Of special interest is also the article by Degon~et~al.~\cite{degon2008}, who presented a mathematical model of Tg production in the thyroid, although the response to RAI was not investigated. {Table \ref{Table_review} summarizes these models, presenting their main characteristics and limitations.}

\begin{table}
\centering
\label{Table_review}
\begin{tabular}{p{0.2\textwidth}p{0.3\textwidth}p{0.3\textwidth}}
\textbf{Author }& \textbf{Summary} & \textbf{Limitation} \\ 
\hline
  Barbolosi~et~al.~\cite{barbolosi2017} & Model of metastatic thyroid cancer response to RAI, including dynamics of Tg. & Phenomenological model, which may limit the potential for individualization. TgAb is not considered. \\ 
\hline
  Traino~et~al.~\cite{traino2006} & DTC mass reduction model in response to RAI. Based on personalized dosimetry calculations. & Extension of the phenomenological Linear--Quadratic formalism to model mass reduction. Neither Tg nor TgAb are considered. \\ 
\hline
  Degon~et~al.~\cite{degon2008} & Normal functioning thyroid model considering TSH, Tg, T3, T4 and TPO.  &  Neither the response to RAI nor Tg are considered. \\ 
\hline
  Pandiyan~et~al.~\cite{pandiyan2018}  & Graves' disease response model to thioamides therapy, modeling the evolution of TSH, T4 and TrAb. & The model does not include response to RAI, and dynamics of Tg and TgAb are not considered.\\ 
\hline
  Langenstein~et~al.~\cite{langenstein2016} & Graves' disease response model to thioamides therapy, modeling  thyroid size, T4 and TrAb. & The model does not include response to RAI, and dynamics of Tg and TgAb are not considered. \\
\hline
 Traino~et~al.~\cite{traino2005} & Graves’ disease mass reduction model, based on personalized RAI dosimetry calculations. & Extension of the phenomenological Linear--Quadratic formalism to model mass reduction. Neither Tg nor TgAb are considered. \\
%\bottomrule
%\end{tabularx}}%
\hline
\end{tabular}
\caption{List of reviewed mathematical models of thyroid/thyroid cancer response.}
\end{table}

In this work, we present a radiobiological model of thyroid response to RAI. We build on the models presented above, while aiming to address some of their limitations. The approach that we follow in this work is more mechanistic than those presented in previous studies~\cite{barbolosi2017,traino2006,traino2005}, and the model specifically includes the dynamics of Tg and TgAb. It consists of a set of differential equations describing the dynamics of different populations of thyroid/tumor cells, and Tg and TgAb concentrations.

The model was fit to published data of patients with metastatic differentiated thyroid cancer treated with several cycles of RAI, and patients with Graves’ disease. Direct intercomparison with previous thyroid response models was, however, not possible, as none of them considered together all the variables included in our work: they either missed the dynamics of Tg or TgAb, or did not model the response to RAI.

In addition, we present a stochastic Markov-like version of the model that can be used to compute tumor control probabilities. We used it to investigate RAI treatment individualization strategies aiming to improve the tumor control rates achieved with this therapy. This latter study highlights the potential importance of such models for the individualization of RAI.

\section{Materials and Methods}

\subsection{The Model}
The response of the thyroid cells to RAI is modeled by extending a previous work of the group that describes the response of a population of cells to a continuous radiation dose rate~\cite{neira2020}. A system of coupled differential equations is used to model the dynamics of different tumor cell populations: undamaged cells, $N(t)$, sub-lethally damaged cells, $N_s(t)$, doomed cells, $N_d(t)$, and dead cells, $N_x(t)$.

The radiation treatment is characterized by a dose rate, $r(t)$, having a time variation that depends on the physical decay of the radioisotope and its biological clearance. Following the rationale behind the classical linear-quadratic model of radiation response~\cite{kellerer1978}, radiation damage can be either lethal ($\sim ar(t)$) or sub-lethal ($\sim br(t)$). Lethally damaged (doomed) cells cannot recover, and will eventually die, moving to the dead compartment with a rate $\gamma$. Sub-lethally damaged cells may recover (we assume a simple exponential repair kinetic term, with characteristic repair rate $\mu$), or may become doomed due to further lethal or sub-lethal damage. Dead cells disappear due to resorption with a resorption\mbox{ rate $\eta$}. 

Cell proliferation is supposed to follow a logistic function, with an exponential rate, $\lambda$, and exponential proliferation moderated according to the total number of cells and a saturation limit, $N_{sat}$.  Experimental evidence shows exponential proliferation slowing down with increasing tumor volumes, and therefore growth models must include some sort of saturation mechanism~\cite{kim2005}. In general, we assume that $N_d$ and $N_s$ cells may still carry proliferative capacity, which can be different from that of undamaged tumor cells.

The system of equations has the following form:

\begin{eqnarray}
\frac{{\rm{d}} N(t)}{{\rm{d}}t} &=& \lambda N(t) [1-N_T(t)/N_{sat}] + \mu N_s(t) - (a+b) r(t) N(t) \label{eq_n1}\\
\frac{{\rm{d}} N_s(t)}{{\rm{d}}t} &=& \lambda_s N_s(t) [1-N_T(t)/N_{sat}] + br(t)N(t)- \mu N_s(t) - (a+b) r(t) N_s(t) \label{eq_n2}\\
\frac{{\rm{d}} N_d(t)}{{\rm{d}}t} &=& \lambda_d N_d(t) [1-N_T(t)/N_{sat}] + (a+b) r(t) N_s(t) + ar(t)N(t) - \gamma N_d(t) \label{eq_n3}\\
\frac{{\rm{d}} N_x(t)}{{\rm{d}}t} &=& \gamma N_d(t) - \eta N_x(t), \label{eq_n4}
\end{eqnarray}
where $N_T(t)=N(t)+N_s(t)+N_d(t)+N_x(t)$ is the total number of cells at time $t$.

Three compartments are included to model the concentration of serum thyroglobulin and antithyroglobulin antibodies in the patient. The free thyroglobulin compartment, $T_f$, stands for the concentration of thyroglobulin not bound to antithyroglobulin antibodies, $A_b$, while the $T_b$ compartment accounts for the concentration of Tg bound to these antibodies:

\begin{eqnarray}
\frac{{\rm{d}} T_f(t)}{{\rm{d}}t} &=& \lambda_{T_g} \frac{N(t)+N_s(t)+N_d(t)}{c+[T_f(t)+T_b(t)]} + z_t N_d(t) - k_e T_f(t) - k_a T_f(t) A_b(t) \label{eq5}\\
\frac{{\rm{d}} T_b(t)}{{\rm{d}}t} &=& k_a T_f(t)A_b(t) - (k_e + k_{eb})T_b(t) \\
\frac{{\rm{d}} A_b(t)}{{\rm{d}}t} &=& z [T_f(t) + T_b(t)] + z_b N_d(t) - k A_b(t) - k_a' T_f(t)A_b(t).
\end{eqnarray}

Patients treated with $^{131}$I for thyroid carcinoma distant metastases or Graves’ disease may achieve complete remission, including undetectable serum Tg levels, months or even years after treatment~\cite{schlumberger1996,becker1982,pacini2001}. This suggests that, after being damaged by radiation, thyroid cells can still produce detectable amounts of Tg for long periods of time, despite having a limited proliferative capacity~\cite{durante2006,baudin2003}. Based on this, we assume that free thyroglobulin is produced by the cells that keep proliferative capacity (healthy, sub-lethally damaged, and doomed cells). The Tg production rate is modulated by the total thyroglobulin level (the parameter $c$ is a modulation constant). This intends to mimic the Hypothalamus--Pituitary--Thyroid feedback mechanisms that regulate the thyroid hormones pools in the body, as previously considered by other thyroid mathematical models~\cite{pandiyan2018,degon2008}.

Moreover, Tg can be released from the colloid reservoir into the bloodstream when thyroid cells are damaged by radiation~\cite{latrofa2018}. To reflect this, a $T_f$ production term is included, proportional to $N_d$, with a characteristic release rate $z_t$. Free Tg is eliminated from blood with a clearance rate $k_e$, and can bind to Tg antibodies (if these are present) with a binding rate $k_a$, moving to a compartment that describes the kinetics of the antibody-bound thyroglobulin serum, $T_b(t)$. This binding also removes TgAb, for which we use the same term with a binding rate $k_a'$. Bound Tg is assumed to show an enhanced metabolic clearance, with rate $k_e+k_{eb}$, induced by TgAb~\cite{latrofa2016,latrofa2018,ricci2020}.

TgAb are intra-follicular antibodies that can bind to immune cells and antigens, not directly linked with thyroid cell destruction per se~\cite{jo2018}. Thyroid disease is associated with Tg structural changes and leakage, leading to TgAb production~\cite{ricci2020,jo2018}. TgAb are mainly produced by lymphocytes infiltrating the gland and, to a lesser extent, by cervical lymph nodes and bone marrow immune cells~\cite{chiovato2003}. Considering this, the TgAb compartment in our model includes two terms of antibody production. The first is proportional to the total serum Tg concentration (through a production rate constant $z$), therefore being related to the cancer thyroid tissue that produces Tg. The second term is related to lymphocyte production, and, therefore, linked to the population of doomed cells that would release antigens, with an associated production constant $z_b$. A natural elimination term is included to model the disappearance of thyroid antibodies observed in patients after thyroidectomy or thyroid tissue destruction by RAI. The previously mentioned binding of Tg to TgAb also results in an effective elimination for these antibodies, as bound TgAb cannot be detected by the immunometric TgAb assays, in which Tg is the agent used to bind the antibodies in the sample~\cite{elecsysDoc}.

\subsection{Modeling Tumor Control Probability---Stochasticity}\label{section 2.2}

In radiobiological modeling, one of the main aims is to model the probability that a given treatment may control the tumor in a population of patients, the so-called Tumor Control Probability (TCP). This is usually mechanistically modeled by assuming the clonogenic cell hypothesis: a tumor will be controlled if all cells with clonogenic capacity in it are killed~\cite{munro1961}. In our model, we associate clonogenic potential to undamaged and sub-lethally damaged cells, $N$ and $N_s$ (doomed $N_d$ cells may still have some proliferative capacity, but they will eventually die and cannot sustain a tumor long-term). Therefore, a given treatment will lead to tumor control if $N(t)+N_s(t)=0$ at some time $t$.

\begin{figure}	
\centering
\includegraphics[width=5.5 cm]{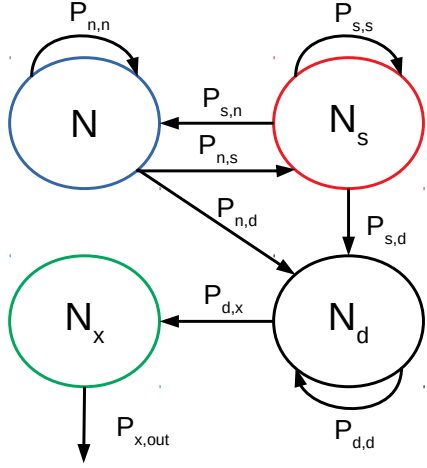}
\caption{Sketch of our stochastic model with transfer probabilities between compartments and flux arrows. $P_{i,j}$ is the probability of a cell in compartment $i$ to go to compartment $j$ (i.e.,~$N_i \rightarrow N_i-1$, $N_j \rightarrow N_j+1$). $P_{i,i}$ is the probability of duplication of a cell in compartment $i$ (i.e.,~$N_i \rightarrow N_i+1$). (\emph{n}, \emph{d}, \emph{s}, \emph{x}) refer to compartments of non-damaged, doomed, sub-lethally damaged, and dead cells, and ``{out}'' is used for cells leaving the system.\label{fig_markov}}
\end{figure}

Notice that our model, as presented in the previous subsection, is continuous and deterministic. In order to simulate tumor control probability, we need a discrete (numbers of cells) stochastic model. To do so, and to be able to model TCP, we have identified terms in the equations as transfer probabilities between compartments, and used them to obtain a stochastic Markov model. The model is sketched in Figure \ref{fig_markov}. Transfer probabilities for each compartment/process are obtained by re-writing the terms in the differential equations as $dN_i/dt = \sum_j f_j(t,N_i, \cdots)N_i(t)$, where $i$ stands for each cell compartment, and each term $f_j$ is a transition probability. For each time step, given by $\Delta t$, the discrete probabilities are given by:

\begin{eqnarray}
P_{n,n} &=& \lambda (1-N_T(t)/N_{sat})\Delta t\\
P_{s,s} &=& \lambda_s (1-N_T(t)/N_{sat})\Delta t\\
P_{d,d} &=& \lambda_d (1-N_T(t)/N_{sat})\Delta t\\
P_{n,s} &=& b r(t)\Delta t\\
P_{n,d} &=& a r(t) \Delta t\\
P_{s,n} &=& \mu \Delta t\\
P_{s,d} &=& (a+b) r(t) \Delta t\\
P_{d,x} &=&  \gamma \Delta t \\
P_{x,out} &=& \eta \Delta t
\end{eqnarray}

$P_{i,j}$ is the probability of a cell in compartment $i$ to go to compartment $j$ (i.e.,~$N_i \rightarrow N_i-1$, $N_j \rightarrow N_j+1$). $P_{i,i}$ is the probability of duplication of a cell in compartment $i$ (i.e.,~$N_i \rightarrow N_i+1$). \emph{n}, \emph{d}, \emph{s}, and \emph{x} refer to compartments of non-damaged, doomed, sub-lethally damaged, and dead cells, respectively, and ``{out}'' is used for cells leaving the system.

In order to speed up computation time (stochastic modeling is more time-consuming), continuous and stochastic modeling have been combined. The stochastic modeling is activated only when numbers of cells fall below a given threshold (10,000 cells). The combination of continuous and discrete modeling seems adequate, as it is much faster than a purely discrete model, and differences between both approaches only appear for low numbers of cells.

\subsection{Fit to Experimental Data: Graves’ Disease and Tumor Response}
\label{data_exp}

There is limited literature providing adequate data to evaluate the capability of our model to fit response to therapy, including the dynamics of cells (mass), Tg, and TgAb. Suitable datasets should include the evolution of thyroid mass/volume, the dynamics of Tg and TgAb after RAI, and the dose to the thyroid/metastases. In a dedicated search, we only found two sets of experimental clinical data, including these variables~\cite{latrofa2018,deKeizer2003}.

The study of Latrofa~et~al.~\cite{latrofa2018} monitored the response to a single administration of 555 MBq RAI in a cohort of 30 patients with Graves’ hyperthyroidism. Thyroid tissue volume and serum Tg and TgAb levels were registered at the time of treatment and every 15 days thereafter, up to 3 months. We used reference cell density values to estimate the corresponding thyroid cell numbers~\cite{delMonte2009}, and average values of absorbed dose per MBq and effective half-life reported by Traino~et~al.~\cite{traino2005}.

We also used the model to fit the experimental data of differentiated thyroid cancer response published by de Keizer~et~al.~\cite{deKeizer2003}. In this study, the authors reported the response of a cohort of patients who suffered $^{131}$I-avid metastatic or recurrent disease after near-total thyroidectomy followed by ablative RAI. Patients received recombinant human thyrotropin (rhTSH) before radioiodine administration of 7400 MBq. Tumor response was assessed by comparing serum Tg levels pre-treatment and 3 months post-treatment. In order to have information about tumor mass evolution and at least three follow-up marker measurements, we focused on the patients who received at least two RAI administrations. This study provided personalized dosimetry information, based on $^{131}$I scintigrams, that allows us to define the dose rate profiles in the lesions, $r(t)$, considering variations in $^{131}$I uptake and biodistribution (effective half-life) during the course of the treatment. These are relevant parameters that importantly affect the outcome, but are rarely available from the literature. Uncertainties of tumor mass, Tg and TgAb levels are not reported in that paper, hence we have associated {ad hoc} uncertainties of 10\% (30\%) of the first measurement point to tumor masses and Tg levels (TgAb levels).

\subsection{Evaluation of Tumor Control Probabilities---Modeling Study}

We have used the stochastic flavour of the model to investigate TCP {in silico} and to formulate hypotheses of optimal treatment strategies. In order to generate a {population} of patients we have taken a set of model parameter values (obtained from fits to experimental data), including uptake and half-life of activities in the tumor, and randomly perturbed them (with normal perturbations with a relative standard deviation of 25\%). With this procedure, we have generated 1000 different sets of parameter values, naively representing a cohort of 1000 patients with individualized responses. 

The TCP for the population (or subgroup of the population) is computed as:

\begin{equation}
TCP= \frac{controls}{size},
\end{equation}
where {controls} refers to the number of simulations (patients) for which control was achieved (as detailed in Section \ref{section 2.2}), and {size} refers to the number of simulations (patients), either in the whole cohort (1000), or in subgroups of that cohort. 

\begin{sidewaysfigure}
\includegraphics[width=\columnwidth]{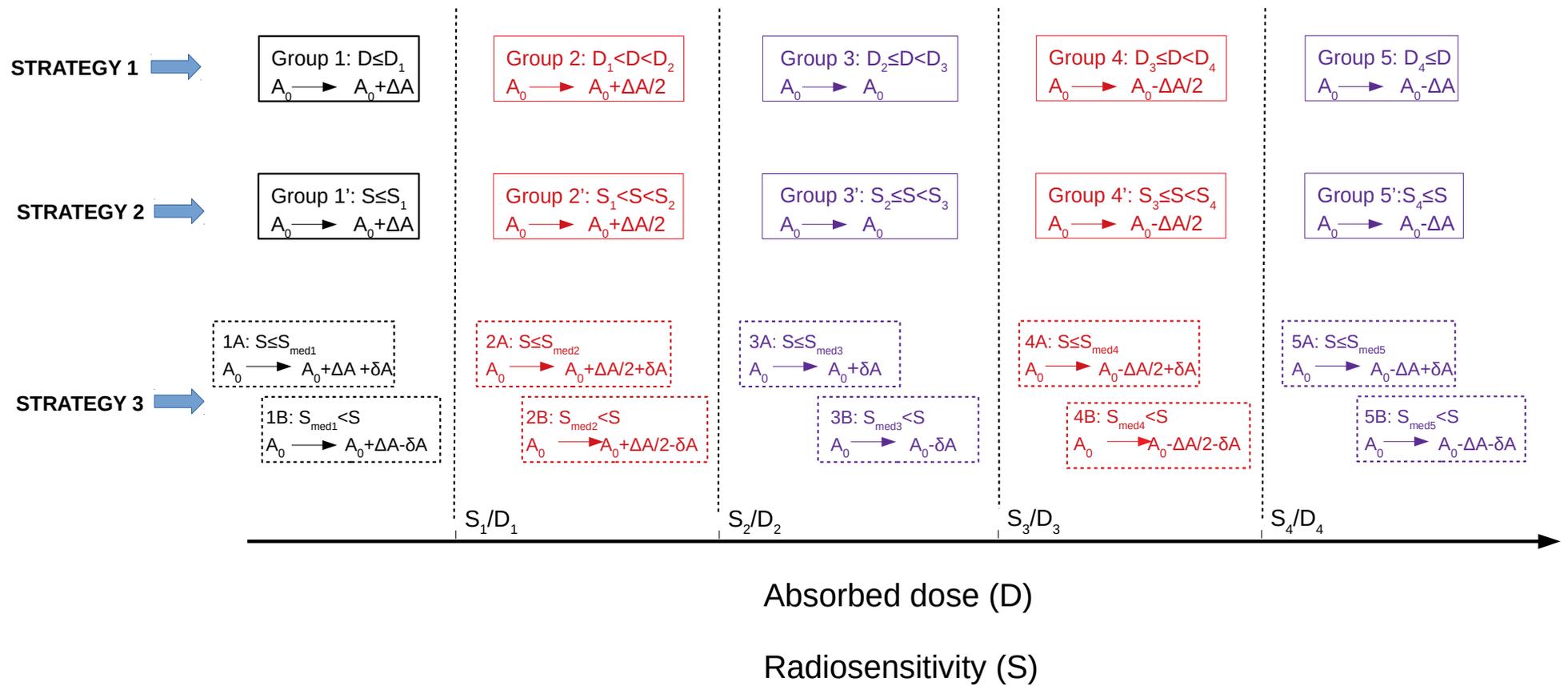}%
\caption{Sketch of stratification and therapy individualization. Three strategies are presented. In  {Strategy 1}, simulated patients are split into 5 groups according to absorbed dose in the tumor. Dose thresholds ($D_1$, $D_2$, $D_3$, $D_4$) are defined to have 1/5 of patients in each group. {Strategy 2} is similar, but patients are stratified into five groups according to their radiosensitivity (characterized through the parameter of accumulation of sub-lethal damage, $b$). Sensitivity thresholds ($S_1$, $S_2$, $S_3$, $S_4$) are defined to have 1/5 of patients in each group. {Strategy 3} is a combination of both previous strategies: the same five groups of {Strategy 1} are created (according to absorbed dose in the tumor), but each of them is further split into two subgroups (A and B) according to the radiosensitivity of tumor cells, as in {Strategy 2}. The threshold to split each group is defined as the median of the group, in order to have 1/2 of patients in subgroup A and subgroup B. For individualization, administration of $^{131}$I activity $A$ is prescribed differently in each group ({Strategy 1} and {Strategy 2}) or subgroup ({Strategy 3}), as a function of  activity perturbations $\Delta A$ and $\delta A$. Individualized activities are defined in such a way to verify that the mean injected activity in the whole cohort does not change, no matter what individualization strategy is used.}
\label{fig_cohort}
\end{sidewaysfigure}

We have investigated whether treatment individualization can lead to important gains of TCP, i.e.,~rather than injecting the same activity $A_0$ to each patient, the injected activity is adjusted according to patient characteristics/model parameters (using parameter values that can be easily measured in the clinic). We have used a single fraction treatment, and we have investigated three hypotheses and individualization strategies, graphically sketched in Figure~\ref{fig_cohort}:

\begin{itemize}

\item {\bf {Individualization strategy 1}} 
---Patients with low absorbed dose in the tumor (either due to low activity uptake and/or fast activity clearance) will have poor tumor control, while patients with high doses in the tumor will have good control. Patients can be stratified according to the absorbed tumor doses associated with the non-individualized standard treatment and the injected activities adjusted for each group. For high-dose patients, if the TCP is already high, a reduction in activity (and therefore dose) might not have a significant effect on tumor control, but could potentially decrease the likelihood of organ/tissue toxicity. On the other hand, for patients receiving low tumor doses, injected activities could be increased to enhance the tumor doses and improve the likelihood of control. According to this reasoning, patients have been stratified into five groups according to absorbed tumor doses, and the injected activities adjusted for each group. In a clinical scenario, one should consider how such changes in injected activities affect the probability of toxicity. Modeling toxicity is beyond the scope of this work, but we have imposed a constraint in the individualization, requiring that the average injected activity in the population does not change (a sort of populational isotoxicity constraint).

\item {\bf {Individualization strategy 2}}---The radiosensitivity of tumor cells will also condition tumor response and tumor control. While radiosensitivity cannot be as easily measured as the radiation dose, there are genetic profiling techniques that can infer the radiosensitivity of cells: this is behind the Genomically Adjusted Radiation Dose (GARD) methodology, aiming at individualizing radiation dose according to radiosensitivity profiling~\cite{ahmed2019,baine2017}. We have investigated a similar approach. Patients have been stratified into five groups, as in Strategy 1, according to their radiosensitivity (the parameter $b$ of sublethal damage, which will serve as a simple surrogate of radiosensitivity, even if radiosensitivity will depend on more parameters). Those in the groups with lower $b$ values (radioresistant) will have the injected activity scaled up, and those in groups with higher $b$ values (radiosensitive) will receive scaled-down activities, as shown in Figure~\ref{fig_cohort}.

\item {\bf {Individualization strategy 3}}---We will also investigate a strategy that is a combination of 1 and 2 above. The dose groups created in {Individualization strategy 1} have been further split into two subgroups, A and B, each containing 50\% of the patients in the group, according to the value of parameter $b$. Those in the subgroup with lower $b$ values (radioresistant) will have the injected activity scaled up, and those in the subgroup with higher $b$ values (radiosensitive) will receive scaled-down activities.
\end{itemize}

\subsection{Model Implementation, Resolution, Identifiability, and Data Fitting}
The model was implemented in Matlab (The Mathworks Inc.) and it was numerically solved by employing an explicit Euler method~\cite{press1987} with a time step $\Delta t$ = 5 min. The same time step was used for the stochastic Markov model. The integration time depends on the problem under study, but can reach several months (for multifraction treatments of DTC). In this situation, only a fraction of the simulated data were saved (typically data are saved every day) in order to prevent unmanageable datasets. Code and data are available from the Harvard Dataverse~\cite{gago2021}, including a more detailed presentation.

A simulated annealing method~\cite{kirkpatrick1984} was implemented to fit the model to experimental data. The function to be minimized, $F$, during the optimization was the weighted sum of quadratic differences between model and experimental data:

\begin{equation}
F=\sum_i \sum_j \frac{(A^{exp}_{i,j} - A^{mod}_{i,j})^2}{w_{i,j} ^2},
\end{equation}
where $A^{exp}$ and $A^{mod}$ refer to experimental and model results, and $w$ to the weights (uncertainties of the experimental data presented in Section~\ref{data_exp}). Indices $i$ and $j$ run over the number of variables (volumes/masses, Tg, TgAb) and time points included in the fit, respectively.

Identifiability is an important property when fitting a model to experimental data and refers to whether it is possible to determine the values of the unknown parameters. We have investigated the identifiability of the model here presented by relying on the DAISY package~\cite{bellu2007}.

\section{Results}

\subsection{Identifiability}

The model is classified as globally identifiable by DAISY, when considering the following observables (and the rationale for considering them): 

\begin{itemize}
\item $r(t)$: the dose rate could be monitored by using, for example, imaging techniques to characterize the dynamics of activity/dose. Formally, $r(t)$ is an input, rather than an~observable.
\item $N(t)+N_s(t)+N_d(t)+N_x(t)$: this is related to the dynamics of tumor volumes, which can be again monitored with imaging techniques.
\item $N(t)+N_s(t)$: in our model, $N$ and $N_s$ carry clonogenic capacity, and the fraction/\linebreak number of clonogenic cells can be obtained by harvesting tumor samples and performing plating experiments.
\item $T_b(t)+T_f(t)$: the dynamics of serum thyroglobulin can be monitored in blood samples. Separating $T_b$ from $T_f$ may be more difficult, therefore we consider the sum of both variables.
\item $A_b(t)$: the dynamics of serum thyroglobulin antibodies can be also monitored in blood~samples.
\end{itemize}

\subsection{Fit to Graves' Disease Data}

\begin{figure}	[H]
%\widefigure
\centering
\includegraphics[width=12 cm]{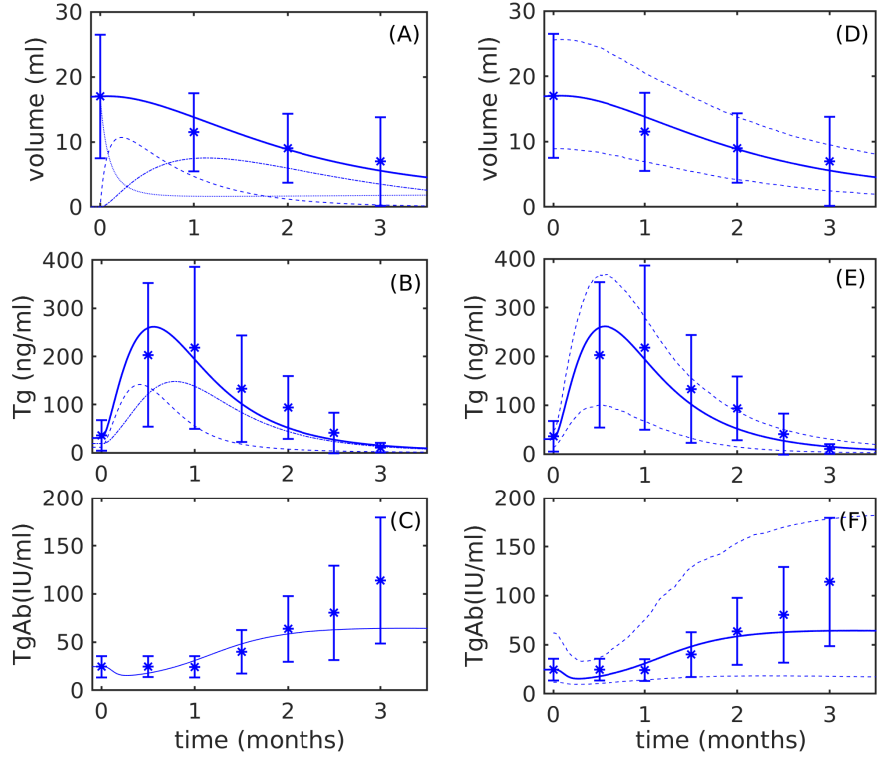}
\caption{Evolution of population-averaged thyroid markers with time after $^{131}$I treatment for Graves' disease. Experimental measurements, asterisks with standard deviation error bars, and best fits obtained with the model presented in this work: (\textbf{A}) thyroid volume: total, thick solid line; and volumes corresponding to undamaged cells, dot line; doomed cells, dashed line; and dead cells, dash-dot line. (\textbf{B}) Tg concentration: total, solid line; free Tg compartment, dashed line; and bound Tg compartment, dash-dot line. (\textbf{C}) TgAb concentration. Panels (\textbf{D}--\textbf{F}) show 68\% confidence intervals for thyroid volume, Tg, and TgAb, respectively (dashed lines), as well as population-averaged fits. In order to simulate a population, best-fitting parameters to the population-averaged data were perturbed with normal distribution with 25\% relative standard deviation (1000 different sets of parameter values to simulate a population of 1000 patients).
\label{fig_graves}}
\end{figure}

In Figure \ref{fig_graves} we present best-fits to population-averaged response to a single administration of 555 MBq RAI measured in a cohort of 30 patients with Graves’ hyperthyroidism. Left panels show fits of the model to volumes (total volumes and fractions corresponding to undamaged, doomed, and dead cells), as well as Tg (total, free, and bound) and TgAb. Best-fitting parameter values are reported in the supplementary materials, Table~\ref{tableSM}. In general, results obtained with the model match fairly well the trends observed in the measurements. Slow-doomed and dead-cell elimination rates, with half-lives $\sim$ 1 month, were required to fit the smooth decrease in thyroid volume with time after treatment. Tg and TgAb levels were adequately reproduced, considering the experimental uncertainties. However, the rise observed in TgAb after month 2 (five-fold increase with respect to the minimum value registered) was much smoother in the model (four-fold change). 

We have also simulated the effect of interpatient variabilities. In order to simulate a population, best-fitting parameters to the population-averaged data were perturbed with a normal distribution with 25\% relative standard deviation (1000 different sets of parameter values to simulate a population of 1000 patients). Confidence intervals reported in Figure \ref{fig_graves} (left panels) are similar to the variability of the experimental data.

\subsection{Fit to Differentiated Thyroid Cancer Response Data}

\begin{figure}	[H]
%\widefigure
\centering
\includegraphics[width=8 cm]{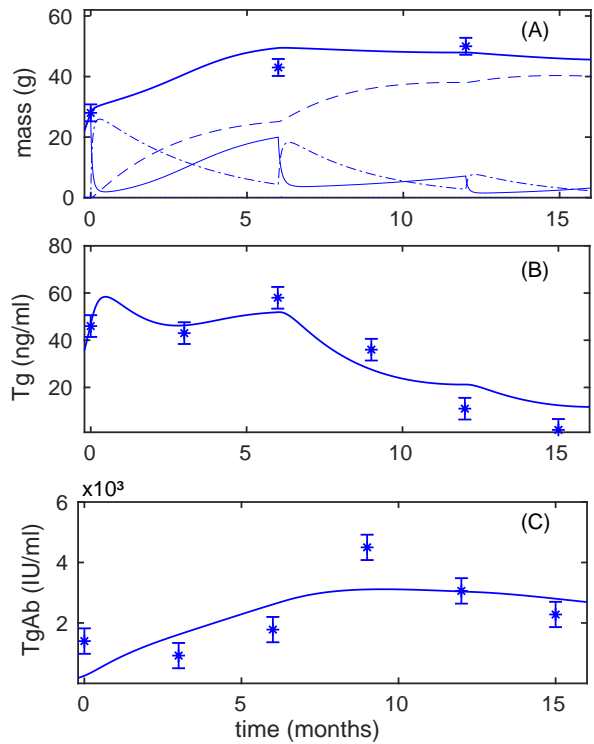}
\caption{Evolution of DTC local recurrence with time for a $^{131}$I treatment consisting of three administrations of 7400 MBq each. Experimental measurements, represented by asterisks with error bars, and best fits obtained with the model. (\textbf{A}) Tumor mass: total indicated by a thick solid line; masses corresponding to undamaged cells by a thin solid line; doomed cells by a dash--dot line; and dead cells by a dashed line. (\textbf{B}) Total Tg concentration. (\textbf{C}) TgAb concentration.
\label{fig_latrofa1}}
\end{figure}

\begin{figure}	[H]
%\widefigure
\centering
\includegraphics[width=12 cm]{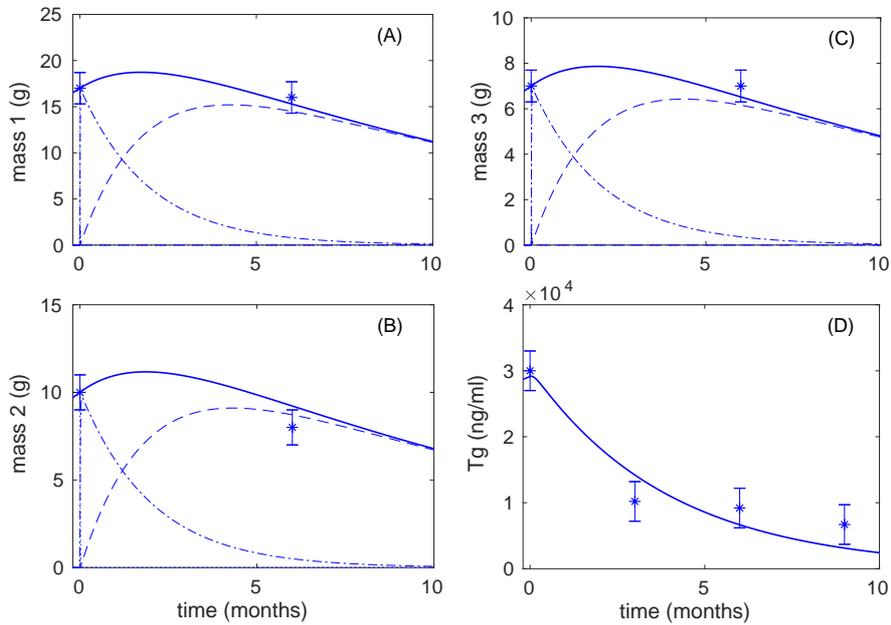}
\caption{Evolution of three DTC lung metastases with time for a $^{131}$I treatment with two administrations of 7400 MBq each. Experimental measurements, asterisks with standard deviation error bars, and best fits obtained with the model. Panels (\textbf{A}--\textbf{C}) show tumor masses corresponding to every lesion: total by a  thick solid line; masses corresponding to doomed cells by a dash-dot line; and dead cells by a dashed line. Panel (\textbf{D}) shows total thyroglobulin concentration.
\label{fig_latrofa2}}
\end{figure}

Figures~\ref{fig_latrofa1} and ~\ref{fig_latrofa2} show best fits of the model to the response of two differentiated thyroid cancer patients treated with RAI after rhTSH stimulation. Best-fitting parameter values are reported in the supplementary materials,  Table~\ref{tableSM}. Patient 1 suffered a local recurrence (one single lesion) and showed detectable TgAb levels (Figure~\ref{fig_latrofa1}). This patient received three $^{131}$I administrations, at months 0, 6, and 12. The tumor mass steadily increased with time after treatment, illustrating the previously reported typical slow response of DTC to RAI. Probably linked to the presence of TgAb, this patient showed rather low Tg levels (panel B), which increased from month 0 to 6, and then decreased to low levels. The model was able to reproduce these trends, although with TgAb changes that are smoother than those experimentally observed (panel C).

Patient 2 presented three metastatic lesions in the lungs, and undetectable TgAb levels. Panels A, B and C in Figure~\ref{fig_latrofa2} show the evolution of the mass with time for every lesion, which remained stable in the follow-up for lesions 1 and 3, and decreased approximately by 20\% at month 6 in lesion 2. Slow-cell elimination rates were again required to fit the evolution of the mass and reproduce the drop in Tg concentration (panel D) associated with the loss of functional, undamaged cells.

\begin{table}[H]
\small
\centering
%%% \tablesize{} %% You can specify the fontsize here, e.g., \tablesize{\footnotesize}. If commented out \small will be used.
%\begin{tabular}{lccccc}
\begin{tabular}{|p{0.3\textwidth}| p{0.073\textwidth}| p{0.14\textwidth}| p{0.11\textwidth}| p{0.12\textwidth}| p{0.12\textwidth}|}

\hline
\textbf{Parameter}	&\textbf{Symbol}	& \textbf{Unit} & \textbf{Graves' hyperthyroidism}	& \textbf{DTC local recurrence } & \textbf{DTC lung metastases} \\
\hline
Proliferation rate of undamaged and sublethally damaged cells*  & $\lambda$, $\lambda_{s}$   & day$^{-1}$ & 2.31E-03 & 0.076 & 5.55E-03 \\ \hline
Proliferation rate of doomed cells &$\lambda_{d}$ & day$^{-1}$& 3.17E-02 & 5.55E-03 & 5.55E-03 \\ \hline
Tumor cell proliferation saturation parameter & N$_{sat}$ &  & 3.40E+09 & 5.14E+09 & 1.10E+10 \\ \hline
Sublethally damaged cells repair rate & $\mu$  & day$^{-1}$ & 8.32 & 8.32 & 8.32 \\ \hline 
Doomed cells death rate & $\gamma$ & day$^{-1}$ & 0.0459 & 0.0118 & 0.0216 \\\hline  
Dead cells resorption rate & $\eta$ & day$^{-1}$ & 0.0239 & 9.50E-04 & 2.76E-03 \\\hline 
Lethal damage radiosensitivity parameter & a & Gy$^{-1}$ & 6.50E-03 & 0.450 & 0.450 \\\hline 
Sublethal damage radiosensitivity parameter & b & Gy$^{-1}$ & 8.84E-21 & 0.450 & 0.236 \\\hline 
Tg production rate & $\lambda_{Tg}$ & (ng/ml)$^{2}\cdot$day$^{-1}$ & 8.82E-07 & 5.84E-05 & 0.0135 \\\hline 
Tg production rate modulation parameter & C & ng/ml & 397 & 1.96E+04 & 36.3 \\\hline 
Colloid to bloodstream Tg release rate & z$_{t}$ & ng/(ml$\cdot$day) & 3.17E-08 & 6.30E-13 & 5.21E-10 \\\hline 
Free Tg metabolic clearance rate & k$_{e}$ & day$^{-1}$ & 0.0693 & 0.0693 & 0.154 \\\hline 
Bound Tg enhanced metabolic clearance rate & k$_{eb}$  & day$^{-1}$ & 0.0693 & 0.0693 & - \\\hline 
Free Tg binding rate to TgAb antibodies & k$_{a}$  & ml/(IU$\cdot$day)  & 9.48E-03 & 0.0226 & - \\\hline 
TgAb binding rate to free Tg & k$_{a}$' & ml/(ng$\cdot$day) & 9.48E-03 & 0.0226  & - \\\hline 
TgAb production rate & z & IU/(ng$\cdot$day) & 0.0874 & 0.487 & - \\\hline 
TgAb production rate (lymphocyte related) & z$_{b}$ & IU/(ml$\cdot$day) & 1.15E-09 & 6.45E-11 & - \\\hline 
TgAb natural elimination rate & k & day$^{-1}$ & 1.64E-10 & 0.00289 & - \\\hline 
\end{tabular}
\caption{ List of parameters associated to the model, symbols used in equations, units and values calculated in the fitting process to: average population Graves' hyperthyroidism data (column 3), and individual DTC response data, one local recurrence and one metastatic patient, columns 4 and 5 respectively. (*) The
same proliferation rate was considered for undamaged and sublethally damaged cells for the sake of simplicity. (**) The same value was used for ka and ka’. \label{tableSM}}
\end{table}

%\clearpage
\subsection{In Silico Treatment Individualization}

In Tables~\ref{table1}--\ref{table3} we show the tumor control probabilities calculated with the Markov model, in a cohort of 1000 simulated patients, for {individualization strategies 1, 2 and 3}, respectively. Statistical uncertainties are not presented for the sake of simplicity, but they can be calculated from the reported TCP values and the number of patients in each group, which, using the binomial distribution, leads to values up to $\pm 5\%$ for 100 patients and TCP$\sim$50\%.

\begin{table} [H]
\centering
%%% \tablesize{} %% You can specify the fontsize here, e.g., \tablesize{\footnotesize}. If commented out \small will be used.
\begin{tabular}{ccccccc}
\hline
\textbf{}	& \textbf{Group 1}	& \textbf{Group 2} & \textbf{Group 3}	& \textbf{Group 4} & \textbf{Group 5} & \textbf{Whole cohort}\\
\hline
$TCP_{ni}$ (\%) & 16 & 68.5 & 89.5 & 99.5 & 99 & 74.5\\
$TCP_{ind1}$ (\%) & 26 & 74 & 91.5 & 98 & 98.5 & 77.6\\
$TCP_{ind2}$ (\%) & 39 & 79.5 & 89 & 95 & 98.5 & 80.2 \\
\hline
\end{tabular}
\caption{Overall and group-specific tumor control probability when the injected activity is not adapted to each group ($TCP_{ni}$), and when it is adapted according to \emph{Strategy 1} with $\Delta A=0.1*A_0$ ($TCP_{ind1}$) or $\Delta A=0.2*A_0$ ($TCP_{ind2}$). The stratification in groups is performed according to absorbed dose in the tumor (see Figure~\ref{fig_cohort}), and activities administered to each group are $A_0+\Delta A$ (Group 1), $A_0+\Delta A$/2 (Group 2), $A_0$ (Group 3), $A_0-\Delta A$/2 (Group 4), $A_0-\Delta A$ (Group 5).\label{table1}}
\end{table}

\begin{table}[H]

%%% \tablesize{} %% You can specify the fontsize here, e.g., \tablesize{\footnotesize}. If commented out \small will be used.
\centering
\begin{tabular}{ccccccc}
\hline
\textbf{}	& \textbf{Group 1'}	& \textbf{Group 2'} & \textbf{Group 3'}	& \textbf{Group 4'} & \textbf{Group 5'} & \textbf{Whole cohort}\\
\hline
$TCP_{ni}$ (\%)   & 48.5 & 71   & 77 & 82.5 & 93.5 & 74.5\\
$TCP_{ind1}$ (\%) & 56 &  73.5   & 77 & 83 & 91.5 & 76.2\\
$TCP_{ind2}$ (\%) & 67.5 & 79.5 & 77.5 & 76.5 & 83.5 & 76.9\\
\hline
\end{tabular}
\caption{Overall and group-specific tumor control probability when the injected activity is not adapted to each group ($TCP_{ni}$), and when it is adapted according to \emph{Strategy 2} with $\Delta A=0.1*A_0$ ($TCP_{ind1}$) or $\Delta A=0.2*A_0$ ($TCP_{ind2}$). The stratification in groups is performed according to radiosensitivity of the tumor, characterized by parameter $b$ (see Figure~\ref{fig_cohort}), and activities administered to each group are $A_0+\Delta A$ (Group 1'), $A_0+\Delta A$/2 (Group 2'), $A_0$ (Group 3'), $A_0-\Delta A$/2 (Group 4'), $A_0-\Delta A$ (Group 5').\label{table2}}
\end{table}

Table~\ref{table1} shows the tumor control probabilities calculated with the Markov model, in a cohort of 1000 simulated patients, for the  {individualization strategy 1} (a stratification based on the absorbed dose in the tumor under the standard treatment). As expected, the TCP is higher in the groups receiving larger doses: when there is no individualization, the TCP in the lowest dosage group (group 1) is 16\% versus 99\% in the highest dosage group (group 5, which corresponds to the 20\% of the cohort who receive the larger doses). Treatment individualization, with maximum activity perturbation of $\pm$10\% and $\pm$20\%, leads to an important TCP increase. In groups 1 and 2, who receive the lower doses with the standard activity, the TCP raises from 16\% and 68.5\% up to 39\% and 79.5\% (in the scenario with maximum activity perturbation of $\pm$20\%). On the other hand, in groups 4 and 5, where the individualization leads to lower injected activities, TCP values decrease slightly, up to 4.5\% lower, yet still very close to 100\%. When evaluated on the whole cohort, {individualization strategy 1} leads to an overall TCP improvement of 3.1\% and 5.7\% for the 10\% and 20\% maximum activity boosts, respectively.

Table~\ref{table2} shows the tumor control probabilities for the \emph{individualization strategy 2} (a stratification based on the radiosensitivity of tumors, characterized by the parameter of sub-lethal damage, $b$). As expected, the TCP is higher in the groups with more radiosensitive tumors: when there is no individualization, the TCP in the least radiosensitive group (group 1') is 48.5\% versus 93.5\% in most radiosensitive group (group 5'). Treatment individualization, with maximum activity perturbation of $\pm$10\% and $\pm$20\%, leads to a slight TCP increase. In groups 1' and 2', the TCP raises from 48.5\% and 71\% up to 67.5\% and 79.5\% (in the scenario with maximum activity perturbation of $\pm$20\%). On the other hand, in groups 4' and 5', where the individualization leads to lower injected activities, TCP values decrease by a $\simeq$ 8\%. When evaluated on the whole cohort, \emph{individualization strategy 2} leads to an overall TCP improvement of 1.7\% and 2.4\% for the 10\% and 20\% maximum activity boosts, respectively. These improvements are lower than those obtained with \emph{strategy 1}, probably because, as mentioned, radiosensitivity has a complex dependence on several parameters, while only parameter $b$ has been considered for this stratification.

Table~\ref{table3} presents the results of \emph{individualization strategy 3}, which further splits the 5 patient groups of strategy 1 into two subgroups each, attending to their radiosensitivity. Row 1 corresponds to non-individualized treatment and reflects the effect of dose andradiosensitivity on TCP: subgroups A comprise the more radioresistant patients, who present lower TCP values than patients in subgroups B. This individualization strategy is able to further improve the control probability of the whole population, improving the TCP of the whole cohort by up to 8\% (versus 5.7\% for \emph{strategy 1}).

\begin{table}[H]
\centering
%%% \tablesize{} %% You can specify the fontsize here, e.g., \tablesize{\footnotesize}. If commented out \small will be used.
\begin{tabular}{cccccccccccc}
\hline
\textbf{}	& \textbf{1A}	& \textbf{1B} & \textbf{2A}	& \textbf{2B} & \textbf{3A} & \textbf{3B} & \textbf{4A}	& \textbf{4B} & \textbf{5A}	& \textbf{5B} & \textbf{WC}\\
\hline
$TCP_{ni}$ (\%)   & 2  & 30 & 40 & 97 & 79 & 100 & 99 & 100 & 98  & 100 & 74.5\\
$TCP_{ind1}$ (\%) & 8 & 46 & 51 & 100 & 82 & 100 & 97 & 100 & 98 & 100 & 78.2\\
$TCP_{ind2}$ (\%) & 22 & 60 & 69 & 99 & 84 & 100 & 94 & 100  & 97  & 100 & 82.5\\
\hline
\end{tabular}
\caption{Overall (whole cohort, WC) and group-specific tumor control probability when the injected activity is not adapted to each group ($TCP_{ni}$), and when it is adapted according to \emph{Strategy 3} with $\Delta A=0.1*A_0$ ($TCP_{ind1}$) or $\Delta A=0.2*A_0$ ($TCP_{ind2}$). The stratification in groups is performed according to absorbed dose in the tumor and radioresistance of tumor cells. Activities administered to each group are $A_0+9\Delta A/8$ (Group 1A), $A_0+7\Delta A/8$ (Group 1B), $A_0+5\Delta A/8$ (Group 2A), $A_0+3\Delta A/8$ (Group 2B), $A_0+\Delta A/8$ (Group 3A), $A_0-\Delta A/8$ (Group 3B), $A_0-3\Delta A/8$ (Group 4A), $A_0-5\Delta A/8$ (Group 4B), $A_0-7\Delta A/8$ (Group 5A), $A_0-9\Delta A/8$ (Group 5B). \label{table3}}
\end{table}

%%%%%%%%%%%%%%%%%%%%%%%%%%%%%%%%%%%%%%%%%%

\section{Discussion}

{In this piece of work, we have presented a mechanistic biomathematical model of thyroid response to RAI. The model includes different populations of thyroid cells with varying degrees of radiation-induced damage, and the dynamics of Tg and TgAb, which are important surrogates of treatment response. The model can adequately fit the response to RAI of DTC and Graves' disease observed in experimental studies, as shown \mbox{in Figures~\ref{fig_graves}--\ref{fig_latrofa2}}.

The ultimate goal of {in silico} biomathematical response models is to assist in the design of optimal therapeutic strategies, including treatment individualization. In this regard, we used a stochastic version of the model to investigate the treatment gain (measured in terms of tumor control probability) that could be achieved by DTC RAI individualization. Administered activities were tailored considering patient-specific dosimetry, as well as radiosensitivity features. In a simulated patient cohort, we have found that individualization with moderate variations of the injected activity may lead to an improvement of the TCP (up to $\simeq$+8\% for the whole cohort, and up to $\simeq$+20\% for specific groups of patients, see Tables~\ref{table1}--\ref{table3}), exemplifying how the model can be used to study novel therapeutic strategies. The model can be easily used to analyze other treatment parameters such as the time between RAI administrations in multiple fraction treatments.

Given the large number of parameters involved in our model, it is convenient to use experimental data to set fixed values for some of them, or at least to establish ranges within which the parameter values would be expected to lie, based on population-averaged values or clinical/observational studies. This is of paramount importance to ensure that the model can fit experimentally observed dynamics with parameters that are biologically sound and to avoid overfitting, and merits further discussion:

\begin{itemize}

\item Radiobiological experiments with tumor cell cultures irradiated at different dose rates are useful to determine radiosensitivity parameters and repair rate ranges~\cite{challeton1997}. These data served to fix the repair half-life to 2 hours, within the range of values reported in Steel~et~al.~\cite{steel1987}, and to set a threshold to the radiosensitivity parameters $a$ and $b$ equal to 0.45. The parameter values associated with tumor cell elimination are more uncertain: the mechanisms by which radiation induces the different modes of cell death, their relative relevance, and variability between cancer cell types are still a matter of debate~\cite{balcer2012}. Thyroid tumors are mostly of epithelial origin~\cite{baloch2008}, which has been associated with low levels of early apoptosis. Mitotic catastrophe, observed to peak five days after irradiation, may contribute more importantly to cell-killing in these tumors~\cite{ruth2000,vrezavcova2011}. Other pathways such as senescence or radiation-induced stimuli of the innate and adaptive immune cells would extend the kinetics of cell death (the time when lethally damaged cells die). Although the relative relevance of these pathways remains uncertain~\cite{wu2017}, the delayed effect observed in the response of thyroid to RAI~\cite{pacini2001,baudin2003} suggests that they may play a significant role. As a result, doomed and dead cell elimination rates are expected to be low, with effective half-lives of the order of several weeks/months. The experimental data used in this work for model fitting confirmed this, as masses/volumes decreased slowly, requiring cell elimination half-lives that ranged from 15 days to values as high as 2 years, for the dead cells of the DTC local recurrence.

\item Regarding thyroid tumor markers, the information available to estimate value ranges for many of the parameters associated with the dynamics of these compounds is very scarce, or non-existent in some cases. The natural elimination of serum Tg from the body is an exception to this. The follow-up of Tg levels after thyroidectomy in non-metastatic DTC patients allows us to characterize this process, with measured Tg half-lives ranging from 3.7 h to 4.3 days in humans~\cite{giovanella2010,feldt1978,hocevar1997}. In our data fitting process, Tg elimination rates tended to take values lower than those clinically measured, and we set a maximum Tg half-life of 10 days.   

\item A relation between serum Tg level and tumor burden is commonly suggested, but there are very few quantitative studies on the subject. Bachelot~et~al. measured serum Tg levels after thyroid hormone withdrawal and tumor mass in patients with follicular and papillary thyroid cancer with lymph node metastases and negative TgAb tests~\cite{bachelot2002}. Patients had previously undergone resection of the thyroid gland and showed low (0.5\%) or no significant uptake in the tumor bed, indicating that these metastases were the only source of serum Tg. Lymph node metastases were localized by a preoperative $^{131}$I whole-body scan and completely removed for accurate mass determination. We used the mass and Tg data reported in this study to estimate an average value for bulk Tg production of our model. The large reported interpatient variability of tumor masses and Tg levels led us to set a range for the Tg production rate ($[4.05\times 10^{-4}, 1.1\times 10^{-3}]$ ng$^2$/(ml$^2$ day$^{-1}$)).

\end{itemize}

Best-fitting parameter values for each set of experimental data are presented in the supplementary materials, Table SM1. It is observed that the cell proliferation rates required to fit the response of Graves' disease to RAI are lower than those obtained when fitting thyroid cancer data, which is an expected result due to the typically high proliferative capacity of tumor cells. Best-fitting parameters indicate that the radiosensitivity of cells in Graves' disease is lower than the radiosensitivity of tumor cells in DTC, which is expected due to the typically higher radiosensitivity of rapidly-proliferating tumor cells.

%%%%%%%%%%%%%%%%%%%%%%%%%%%%%%%%%%%%%%%%%%
\section{Conclusions}

In this work, we have presented a mathematical model of thyroid response to radiotherapy. The model includes several populations of cells, undamaged (viable), sub-lethally damaged, doomed (non-repairable), and dead cells, as well as two tumor markers, the serum concentration of thyroglobulin and antithyroglobulin autoantibodies. These two substances are relevant in the management of thyroid cancer patients, being used as surrogates of treatment response in the follow-up after RAI, to evaluate whether tumor control has been achieved or make the decision to deliver further treatment. The model considers several processes such as cell proliferation, sub-lethal and lethal radiation damage, repair, elimination of dead cells, and the release, interaction and elimination of Tg and TgAb. To our knowledge, no previous radiotherapy thyroid response model has considered these two thyroid tumor markers together. Previous response models are restricted to the tumor mass~\cite{traino2006,traino2005}, consider mass and Tg~\cite{barbolosi2017}, or other antibodies relevant for Graves' hyperthyroidism~\cite{pandiyan2018}. However, the interaction of Tg and TgAb is an important feature that is worth to be modeled, as a good understanding of it may contribute to a better understanding of treatment response and improve treatment design. 

%The model is implemented for a continuous dose rate scenario, typical of molecular radiotherapy, which is the radiotherapy strategy most extended to treat thyroid disease, but faster irradiation can be easily simulated.}

Models like this can assist in the interpretation of clinical data, being useful to interpret patient follow-up data{, for example, they may help to determine the presence of subclinical burden from Tg/TgAb dynamics, or when to deliver additional treatment in multi-fraction therapies.} A more ambitious role for this type of model is to assist in the design of optimal therapeutic strategies. Aiming precisely at this objective, we have developed a stochastic version of our model that can be used to compute tumor control probabilities, and have used it to investigate {in silico} treatment individualization according to patient-specific dose/radiosensitivity features.

Among the most significant findings of our work we highlight that (a) the model adequately fits the response to RAI of DTC and Graves' disease observed in experimental studies, and we hypothesize that (b) treatment individualization involving moderate variations of the injected activity can lead to important improvements of TCP.

The model has limitations though, which must be considered for clinical application. In particular, the model contains multiple fitting parameters, and parameter values should be taken with caution, as mass and dosimetry uncertainties strongly affect the values associated with radiosensitivity, interpatient variability is large and there is scarce literature about TgAb-associated parameters. We have not performed a direct intercomparison of fits obtained with our model and previous models (discussed in Section~\ref{sec_intro}), as none of them considered all the variables included in our work and, therefore, a direct comparison was not possible.

Molecular radiotherapy, including RAI, is undergoing an important expansion, due to the good clinical results achieved. Further validation of the presented model will be possible once more RAI response data (including dose, thyroid mass/volume, and Tg and TgAb dynamics) become available. In addition, the same modeling methodology could be of interest to investigate the dynamics of different tumors treated with different molecular radiotherapies. Equations (\ref{eq_n1})--(\ref{eq_n4}), modeling tumor response, could be of direct application to different tumors, while new equations should be developed to account for the dynamics of relevant biomarkers.\\

%%%%%%%%%%%%%%%%%%%%%%%%%%%%%%%%%%%%%%%%%%
\medskip

\textbf{Data Availability Statement:}
Data and code supporting the reported results are available from https://doi.org/10.7910/DVN/PBLNH1.\\

\medskip

\textbf{Funding: }This project has received funding from the Instituto de Salud Carlos III (PI17/01428 grant, FEDER co-funding). This project has received funding from the European Union’s Horizon 2020 research and innovation programme under the Marie Skłodowska-Curie grant agreement No 839135. This project has received funding from FEDER/Ministerio de Ciencia, Innovaci\'on y Universidades---Agencia Estatal de Investigaci\'on, under grant MTM2017-84446-C2-2-R.}

\bibliographystyle{ieeetr}
\bibliography{references}

\begin{thebibliography}{10}

\bibitem{maxon1983}
H.~R. Maxon, S.~R. Thomas, V.~S. Hertzberg, J.~G. Kereiakes, I.-W. Chen, M.~I.
  Sperling, and E.~L. Saenger, ``Relation between effective radiation dose and
  outcome of radioiodine therapy for thyroid cancer,'' {\em New England Journal
  of Medicine}, vol.~309, no.~16, pp.~937--941, 1983.

\bibitem{jentzen2014}
W.~Jentzen, J.~Hoppenbrouwers, P.~van Leeuwen, D.~van~der Velden, R.~van~de
  Kolk, T.~D. Poeppel, J.~Nagarajah, W.~Brandau, A.~Bockisch, and
  S.~Rosenbaum-Krumme, ``Assessment of lesion response in the initial
  radioiodine treatment of differentiated thyroid cancer using 124i pet
  imaging,'' {\em Journal of Nuclear Medicine}, vol.~55, no.~11,
  pp.~1759--1765, 2014.

\bibitem{wierts2016}
R.~Wierts, B.~Brans, B.~Havekes, G.~J. Kemerink, S.~G. Halders, N.~N. Schaper,
  W.~H. Backes, F.~M. Mottaghy, and W.~Jentzen, ``Dose--response relationship
  in differentiated thyroid cancer patients undergoing radioiodine treatment
  assessed by means of 124i pet/ct,'' {\em Journal of Nuclear Medicine},
  vol.~57, no.~7, pp.~1027--1032, 2016.

\bibitem{schlumberger1996}
M.~Schlumberger, C.~Challeton, F.~De~Vathaire, J.-P. Travagli, {\em et~al.},
  ``Radioactive iodine treatment and external radiotherapy for lung and bone
  metastases from thyroid carcinoma,'' {\em The Journal of Nuclear Medicine},
  vol.~37, no.~4, p.~598, 1996.

\bibitem{wang2017}
R.~Wang, Y.~Zhang, J.~Tan, G.~Zhang, R.~Zhang, W.~Zheng, and Y.~He, ``Analysis
  of radioiodine therapy and prognostic factors of differentiated thyroid
  cancer patients with pulmonary metastasis: an 8-year retrospective study,''
  {\em Medicine}, vol.~96, no.~19, 2017.

\bibitem{gleisner2017}
K.~S. Gleisner, E.~Spezi, M.~Aldridge, K.~Bacher, B.~Brans, C.~Chiesa,
  F.~Cicone, C.~Kobe, M.~Konijnenberg, P.~Minguez~Gabina, M.~Paphiti,
  M.~Sandstrom, P.~Solny, J.~Tipping, M.~Wissmeyer, G.~Flux, and C.~Stokke,
  ``Treatment planning for molecular radiotherapy: potential and prospects.
  internal dosimetry task force report,'' 2017.

\bibitem{indrasena2017}
B.~S.~H. Indrasena, ``Use of thyroglobulin as a tumour marker,'' {\em World
  Journal of Biological Chemistry}, vol.~8, no.~1, p.~81, 2017.

\bibitem{giovanella2010}
L.~Giovanella, L.~Ceriani, and M.~Maffioli, ``Postsurgery serum thyroglobulin
  disappearance kinetic in patients with differentiated thyroid carcinoma,''
  {\em Head \& neck}, vol.~32, no.~5, pp.~568--571, 2010.

\bibitem{pelttari2010}
H.~Pelttari, M.~J. V{\"a}lim{\"a}ki, E.~L{\"o}yttyniemi, and
  C.~Schalin-J{\"a}ntti, ``Post-ablative serum thyroglobulin is an independent
  predictor of recurrence in low-risk differentiated thyroid carcinoma: a
  16-year follow-up study,'' {\em European Journal of Endocrinology}, vol.~163,
  no.~5, p.~757, 2010.

\bibitem{piccardo2013}
A.~Piccardo, F.~Arecco, M.~Puntoni, L.~Foppiani, M.~Cabria, S.~Corvisieri,
  A.~Arlandini, V.~Altrinetti, R.~Bandelloni, and F.~Orlandi, ``Focus on
  high-risk dtc patients: high postoperative serum thyroglobulin level is a
  strong predictor of disease persistence and is associated to progression-free
  survival and overall survival,'' {\em Clinical Nuclear Medicine}, vol.~38,
  no.~1, pp.~18--24, 2013.

\bibitem{haugen2016}
B.~R. Haugen, E.~K. Alexander, K.~C. Bible, G.~M. Doherty, S.~J. Mandel, Y.~E.
  Nikiforov, F.~Pacini, G.~W. Randolph, A.~M. Sawka, M.~Schlumberger, {\em
  et~al.}, ``2015 american thyroid association management guidelines for adult
  patients with thyroid nodules and differentiated thyroid cancer: the american
  thyroid association guidelines task force on thyroid nodules and
  differentiated thyroid cancer,'' {\em Thyroid}, vol.~26, no.~1, pp.~1--133,
  2016.

\bibitem{brendel1990}
A.~Brendel, B.~Lambert, M.~Guyot, R.~Jeandot, H.~Dubourg, P.~Roger,
  S.~Wynchauk, G.~Manciet, and G.~Lefort, ``Low levels of serum thyroglobulin
  after withdrawal of thyroid suppression therapy in the follow up of
  differentiated thyroid carcinoma,'' {\em European Journal of Nuclear Medicine
  and Molecular Imaging}, vol.~16, no.~1, pp.~35--38, 1990.

\bibitem{spencer1998}
C.~Spencer, M.~Takeuchi, M.~Kazarosyan, C.~Wang, R.~Guttler, P.~Singer,
  S.~Fatemi, J.~LoPresti, and J.~Nicoloff, ``Serum thyroglobulin
  autoantibodies: prevalence, influence on serum thyroglobulin measurement, and
  prognostic significance in patients with differentiated thyroid carcinoma,''
  {\em The Journal of Clinical Endocrinology \& Metabolism}, vol.~83, no.~4,
  pp.~1121--1127, 1998.

\bibitem{grebe2009}
S.~K. Grebe, ``Diagnosis and management of thyroid carcinoma: a focus on serum
  thyroglobulin,'' {\em Expert Review of Endocrinology \& Metabolism}, vol.~4,
  no.~1, pp.~25--43, 2009.

\bibitem{netzel2015}
B.~C. Netzel, S.~K. Grebe, B.~G. Carranza~Leon, M.~R. Castro, P.~M. Clark,
  A.~N. Hoofnagle, C.~A. Spencer, A.~F. Turcu, and A.~Algeciras-Schimnich,
  ``Thyroglobulin (tg) testing revisited: Tg assays, tgab assays, and
  correlation of results with clinical outcomes,'' {\em The Journal of Clinical
  Endocrinology \& Metabolism}, vol.~100, no.~8, pp.~E1074--E1083, 2015.

\bibitem{spencer2014}
C.~Spencer, I.~Petrovic, S.~Fatemi, and J.~LoPresti, ``Serum thyroglobulin (tg)
  monitoring of patients with differentiated thyroid cancer using sensitive
  (second-generation) immunometric assays can be disrupted by false-negative
  and false-positive serum thyroglobulin autoantibody misclassifications,''
  {\em The Journal of Clinical Endocrinology \& Metabolism}, vol.~99, no.~12,
  pp.~4589--4599, 2014.

\bibitem{latrofa2016}
F.~Latrofa, D.~Ricci, E.~Sisti, P.~Piaggi, C.~Nencetti, M.~Marino, and
  P.~Vitti, ``Significance of low levels of thyroglobulin autoantibodies
  associated with undetectable thyroglobulin after thyroidectomy for
  differentiated thyroid carcinoma,'' {\em Thyroid}, vol.~26, no.~6,
  pp.~798--806, 2016.

\bibitem{latrofa2018}
F.~Latrofa, D.~Ricci, S.~Bottai, F.~Brozzi, L.~Chiovato, P.~Piaggi,
  M.~Marin{\`o}, and P.~Vitti, ``Effect of thyroglobulin autoantibodies on the
  metabolic clearance of serum thyroglobulin,'' {\em Thyroid}, vol.~28, no.~3,
  pp.~288--294, 2018.

\bibitem{ricci2020}
D.~Ricci, A.~Brancatella, M.~Marin{\`o}, M.~Rotondi, L.~Chiovato, P.~Vitti, and
  F.~Latrofa, ``The detection of serum igms to thyroglobulin in subacute
  thyroiditis suggests a protective role of igms in thyroid autoimmunity,''
  {\em The Journal of Clinical Endocrinology \& Metabolism}, vol.~105, no.~6,
  pp.~e2261--e2270, 2020.

\bibitem{jo2018}
K.~Jo and D.-J. Lim, ``Clinical implications of anti-thyroglobulin antibody
  measurement before surgery in thyroid cancer,'' {\em The Korean Journal of
  Internal Medicine}, vol.~33, no.~6, p.~1050, 2018.

\bibitem{sinclair2006}
D.~Sinclair, ``Thyroid antibodies: which, why, when and who?,'' {\em Expert
  Review of Clinical Immunology}, vol.~2, no.~5, pp.~665--669, 2006.

\bibitem{lassmann2010}
M.~Lassmann, C.~Reiners, and M.~Luster, ``Dosimetry and thyroid cancer: the
  individual dosage of radioiodine,'' {\em Endocrine-Related Cancer}, vol.~17,
  no.~3, pp.~R161--R172, 2010.

\bibitem{barbolosi2017}
D.~Barbolosi, I.~Summer, C.~Meille, R.~Serre, A.~Kelly, S.~Zerdoud,
  C.~Bournaud, C.~Schvartz, M.~Toubeau, M.-E. Toubert, {\em et~al.}, ``Modeling
  therapeutic response to radioiodine in metastatic thyroid cancer: a
  proof-of-concept study for individualized medicine,'' {\em Oncotarget},
  vol.~8, no.~24, p.~39167, 2017.

\bibitem{traino2006}
A.~Traino and F.~Di~Martino, ``A dosimetric algorithm for patient-specific 131i
  therapy of thyroid cancer based on a prescribed target-mass reduction,'' {\em
  Physics in Medicine \& Biology}, vol.~51, no.~24, p.~6449, 2006.

\bibitem{pandiyan2018}
B.~Pandiyan, S.~J. Merrill, F.~Di~Bari, A.~Antonelli, and S.~Benvenga, ``A
  patient-specific treatment model for graves’ hyperthyroidism,'' {\em
  Theoretical Biology and Medical Modelling}, vol.~15, no.~1, pp.~1--25, 2018.

\bibitem{langenstein2016}
C.~Langenstein, D.~Schork, K.~Badenhoop, and E.~Herrmann, ``Relapse prediction
  in graves disease: towards mathematical modeling of clinical, immune and
  genetic markers,'' {\em Reviews in Endocrine and Metabolic Disorders},
  vol.~17, no.~4, pp.~571--581, 2016.

\bibitem{traino2005}
A.~C. Traino, F.~Di~Martino, M.~Grosso, F.~Monzani, A.~Dardano, N.~Caraccio,
  G.~Mariani, and M.~Lazzeri, ``A predictive mathematical model for the
  calculation of the final mass of graves' disease thyroids treated with
  131i,'' {\em Physics in Medicine \& Biology}, vol.~50, no.~9, p.~2181, 2005.

\bibitem{degon2008}
M.~Degon, S.~R. Chipkin, C.~Hollot, R.~T. Zoeller, and Y.~Chait, ``A
  computational model of the human thyroid,'' {\em Mathematical Biosciences},
  vol.~212, no.~1, pp.~22--53, 2008.

\bibitem{neira2020}
S.~Neira, A.~Gago-Arias, J.~Guiu-Souto, and J.~Pardo-Montero, ``A kinetic model
  of continuous radiation damage to populations of cells: comparison to the lq
  model and application to molecular radiotherapy,'' {\em Physics in Medicine
  \& Biology}, vol.~65, no.~24, p.~245015, 2020.

\bibitem{kellerer1978}
A.~M. Kellerer and H.~H. Rossi, ``A generalized formulation of dual radiation
  action,'' {\em Radiation Research}, vol.~75, no.~3, pp.~471--488, 1978.

\bibitem{kim2005}
J.~Kim and I.~Tannock, ``Repopulation of cancer cells during therapy: an
  important cause of treatment failure,'' {\em Nature Reviews Cancer}, vol.~5,
  pp.~516--525, 2005.

\bibitem{becker1982}
D.~Becker and J.~Hurley, ``Current status of radioiodine (131 i) treatment of
  hyperthyroidism,'' in {\em Nuclear medicine annual 1982}, 1982.

\bibitem{pacini2001}
F.~Pacini, L.~Agate, R.~Elisei, M.~Capezzone, C.~Ceccarelli, F.~Lippi,
  E.~Molinaro, and A.~Pinchera, ``Outcome of differentiated thyroid cancer with
  detectable serum tg and negative diagnostic 131i whole body scan: comparison
  of patients treated with high 131i activities versus untreated patients,''
  {\em The Journal of Clinical Endocrinology \& Metabolism}, vol.~86, no.~9,
  pp.~4092--4097, 2001.

\bibitem{durante2006}
C.~Durante, N.~Haddy, E.~Baudin, S.~Leboulleux, D.~Hartl, J.~Travagli,
  B.~Caillou, M.~Ricard, J.~Lumbroso, F.~De~Vathaire, {\em et~al.}, ``Long-term
  outcome of 444 patients with distant metastases from papillary and follicular
  thyroid carcinoma: benefits and limits of radioiodine therapy,'' {\em The
  Journal of Clinical Endocrinology \& Metabolism}, vol.~91, no.~8,
  pp.~2892--2899, 2006.

\bibitem{baudin2003}
E.~Baudin, C.~D. Cao, A.~Cailleux, S.~Leboulleux, J.~Travagli, and
  M.~Schlumberger, ``Positive predictive value of serum thyroglobulin levels,
  measured during the first year of follow-up after thyroid hormone withdrawal,
  in thyroid cancer patients,'' {\em The Journal of Clinical Endocrinology \&
  Metabolism}, vol.~88, no.~3, pp.~1107--1111, 2003.

\bibitem{chiovato2003}
L.~Chiovato, F.~Latrofa, L.~E. Braverman, F.~Pacini, M.~Capezzone,
  L.~Masserini, L.~Grasso, and A.~Pinchera, ``Disappearance of humoral thyroid
  autoimmunity after complete removal of thyroid antigens,'' {\em Annals of
  Internal Medicine}, vol.~139, no.~5\_Part\_1, pp.~346--351, 2003.

\bibitem{elecsysDoc}
``Elecsysanti-tg,'' {\em RocheDiagnostics}, vol.~V6.0, 2018-07.
\newblock \newline\url{
  http://labogids.sintmaria.be/sites/default/files/files/anti-tg_2018-07_v6.pdf}.

\bibitem{munro1961}
T.~Munro and C.~Gilbert, ``The relation between tumour lethal doses and the
  radiosensitivity of tumour cells,'' {\em The British Journal of Radiology},
  vol.~34, no.~400, pp.~246--251, 1961.

\bibitem{deKeizer2003}
B.~de~Keizer, B.~Brans, A.~Hoekstra, P.~M. Zelissen, H.~P. Koppeschaar, C.~J.
  Lips, P.~P. van Rijk, R.~A. Dierckx, and J.~M. de~Klerk, ``Tumour dosimetry
  and response in patients with metastatic differentiated thyroid cancer using
  recombinant human thyrotropin before radioiodine therapy,'' {\em European
  Journal of Nuclear Medicine and Molecular Imaging}, vol.~30, no.~3,
  pp.~367--373, 2003.

\bibitem{delMonte2009}
U.~Del~Monte, ``Does the cell number 109 still really fit one gram of tumor
  tissue?,'' {\em Cell Cycle}, vol.~8, no.~3, pp.~505--506, 2009.

\bibitem{ahmed2019}
K.~A. Ahmed, C.~L. Liveringhouse, M.~N. Mills, N.~B. Figura, G.~D. Grass, I.~R.
  Washington, E.~E. Harris, B.~J. Czerniecki, P.~W. Blumencranz, S.~A.
  Eschrich, {\em et~al.}, ``Utilizing the genomically adjusted radiation dose
  (gard) to personalize adjuvant radiotherapy in triple negative breast cancer
  management,'' {\em EBioMedicine}, vol.~47, pp.~163--169, 2019.

\bibitem{baine2017}
M.~J. Baine and C.~Lin, ``Genome-based modeling for adjusting radiotherapy dose
  (gard)-a significant step toward the future of personalized radiation
  therapy,'' {\em Translational Cancer Research}, vol.~6, no.~Suppl 2, p.~S418,
  2017.

\bibitem{press1987}
W.~Press, S.~Teukolsky, W.~Vetterling, and B.~Flannery, {\em Numerical recipes:
  the art of scientific computing}.
\newblock Cambridge University Press, 2007.

\bibitem{gago2021}
A.~Gago-Arias, S.~Neira, F.~Terragni, and J.~Pardo-Montero, ``Replication data
  for "a model of thyroid disease response to radiotherapy”.'' Harvard
  Dataverse, 2021.

\bibitem{kirkpatrick1984}
S.~Kirkpatrick, ``Optimization by simulated annealing: Quantitative studies,''
  {\em Journal of Statistical Physics}, vol.~34, no.~5, pp.~975--986, 1984.

\bibitem{bellu2007}
G.~Bellu, M.~P. Saccomani, S.~Audoly, and L.~D’Angi{\`o}, ``Daisy: A new
  software tool to test global identifiability of biological and physiological
  systems,'' {\em Computer Methods and Programs in Biomedicine}, vol.~88,
  no.~1, pp.~52--61, 2007.

\bibitem{challeton1997}
C.~Challeton, F.~Branea, M.~Schlumberger, N.~Gaillard, F.~de~Vathaire,
  C.~Badie, P.~Antonini, and C.~Parmentier, ``Characterization and
  radiosensitivity at high or low dose rate of four cell lines derived from
  human thyroid tumors,'' {\em International Journal of Radiation Oncology*
  Biology* Physics}, vol.~37, no.~1, pp.~163--169, 1997.

\bibitem{steel1987}
G.~G. Steel, J.~M. Deacon, G.~M. Duchesne, A.~Horwich, L.~R. Kelland, and J.~H.
  Peacock, ``The dose-rate effect in human tumour cells,'' {\em Radiotherapy
  and Oncology}, vol.~9, no.~4, pp.~299--310, 1987.

\bibitem{balcer2012}
E.~Balcer-Kubiczek, ``Apoptosis in radiation therapy: a double-edged sword,''
  {\em Experimental Oncology}, 2012.

\bibitem{baloch2008}
Z.~W. Baloch and V.~A. LiVolsi, ``Unusual tumors of the thyroid gland,'' {\em
  Endocrinology and Metabolism Clinics of North America}, vol.~37, no.~2,
  pp.~297--310, 2008.

\bibitem{ruth2000}
A.~C. Ruth and I.~B. Roninson, ``Effects of the multidrug transporter
  p-glycoprotein on cellular responses to ionizing radiation,'' {\em Cancer
  Research}, vol.~60, no.~10, pp.~2576--2578, 2000.

\bibitem{vrezavcova2011}
M.~{\v{R}}ez{\'a}{\v{c}}ov{\'a}, G.~Rudolfov{\'a}, A.~Tich{\`y},
  A.~Ba{\v{c}}{\'\i}kov{\'a}, D.~Mutn{\'a}, R.~Havelek, J.~V{\'a}vrov{\'a},
  K.~Odr{\'a}{\v{z}}ka, E.~Luk{\'a}{\v{s}}ov{\'a}, and S.~Kozubek,
  ``Accumulation of dna damage and cell death after fractionated irradiation,''
  {\em Radiation Research}, vol.~175, no.~6, pp.~708--718, 2011.

\bibitem{wu2017}
Q.~Wu, A.~Allouch, I.~Martins, C.~Brenner, N.~Modjtahedi, E.~Deutsch, and J.-L.
  Perfettini, ``Modulating both tumor cell death and innate immunity is
  essential for improving radiation therapy effectiveness,'' {\em Frontiers in
  Immunology}, vol.~8, p.~613, 2017.

\bibitem{feldt1978}
U.~Feldt-Rasmussen, P.~H. Petersen, H.~Nielsen, J.~Date, and C.~Madsen,
  ``Thyroglobulin of varying molecular sizes with different disappearance rates
  in plasma following subtotal thyroidectomy,'' {\em Clinical Endocrinology},
  vol.~9, no.~3, pp.~205--214, 1978.

\bibitem{hocevar1997}
M.~Hocevar, M.~Auersperg, and L.~Stanovnik, ``The dynamics of serum
  thyroglobulin elimination from the body after thyroid surgery,'' {\em
  European Journal of Surgical Oncology (EJSO)}, vol.~23, no.~3, pp.~208--210,
  1997.

\bibitem{bachelot2002}
A.~Bachelot, A.~F. Cailleux, M.~Klain, E.~Baudin, M.~Ricard, N.~Bellon,
  B.~Caillou, J.~P. Travagli, and M.~Schlumberger, ``Relationship between tumor
  burden and serum thyroglobulin level in patients with papillary and
  follicular thyroid carcinoma,'' {\em Thyroid}, vol.~12, no.~8, pp.~707--711,
  2002.

\end{thebibliography}

\end{document}